\documentstyle[aps,prb,twocolumn,epsf]{revtex}
\begin{document}
\twocolumn[\hsize\textwidth\columnwidth\hsize\csname
@twocolumnfalse\endcsname

\title{Extension of the Brinkman-Rice picture and the Mott transition$^{\ast}$}
\author{Hyun-Tak Kim$^{\ast\ast}$}
\address{Telecom. Basic Research Lab., ETRI, Taejon 305-350, Korea}
\maketitle{}

\begin{abstract}
In order to explain the metal-Mott-insulator transition, the
Brinkman-Rice (BR) picture is extended. In   the case  of less
than one  as well as one electron per atom, the on-site Coulomb
repulsion is
 given by $U={\kappa}{\rho}^2U_c$ by averaging the electron charge
  per atom over all atomic sites, where ${\kappa}$ is the correlation
   strength of $U$, ${\rho}$ is the band filling
factor,  and $U_c$ is the  critical on-site Coulomb energy. The
effective mass of a quasiparticle is found to be $\frac
{m^{\ast}}{m}=\frac {1}{1-{\kappa}^2{\rho}^4}$ for
$0<{\kappa}{\rho}^2<1$ and seems to follow the heat capacity data
of Sr$_{1-x}$La$_x$TiO$_3$ and YBa$_2$Cu$_3$O$_{7-\delta}$ at
${\kappa}=1$ and $0<{\kappa}{\rho}^2<1$. The Mott transition of
the first order occurs at ${\kappa}{\rho}^2=1$  and  a band-type
metal-insulator transition  takes  place at ${\kappa}{\rho}^2=0$.
This Mott transition is compared with that in the $d=\infty$
Hubbard model.
\\ \\ PACS number(s): 71.27.+a, 71.30.+h,
74.20.Mn, 74.20.Fg \\ \\
\end{abstract}
]
\narrowtext Although several  theoretical  studies have been
performed to reveal the mechanism of the metal-insulator
transition (MIT),$^{1-4}$ the Mott MIT (called "Mott transition")
in 3$d$ transition-metal oxides including strongly correlated
high-$T_c$ superconductors still remains to be clarified.$^{5,6}$
In particular, the effective mass near the Mott transition has
attracted special attention  because it is related to both the
mechanism of the MIT and the two-dimensional density of states
(2D-DOS). The latter offers a clue to explain  the mechanism of
high-$T_c$ superconductivity.

In this paper,  we build  an extended BR  picture with  a
generalized  effective mass depending on the carrier density,
which reduces to the BR picture in  the case of one electron   per
atom,  and     apply it to experimental data of
Sr$_{1-x}$La$_x$TiO$_3$ and YBCO.

In strongly correlated  metals with one  electron per  atom on a
$d=\infty$ simple cubic lattice, the on-site Coulomb repulsion $U$
is always given. However, in real metallic crystals, in which the
number of electrons  $n$ is less than the number of atoms $m$, $U$
is determined not uniquely but instead by probability,  because
the electronic band structure between sites differs and this
system does not transform from real-space to $K$-space. Therefore,
when charges on a site are averaged over all atomic sites, $U$ can
be defined, such as in the case of one electron per atom.

In the case when $n=m$, the existence probability
($P=n/m=\rho=\rho_{\uparrow}+\rho_{\downarrow}$, "the band filling
factor") of electrons on nearest neighbour sites is one. The
on-site Coulomb interaction of  two electrons in the conduction
band is given by $U=U'{\equiv}{\langle\frac{e^2}{r}\rangle}$. In
the case when $n<m$, $P<1$ and the on-site charge is
$e^{\prime}=eP$ ~(or $e^{\prime}/e=P$), when averaged over sites.
Then, the Coulomb energy is given by $U=P^2U^{\prime}$. $U'$  does
not necessarily agree with the critical   value
$U_c=8{\vert}{\bar{\epsilon}}{\vert}$ of the interaction in  the
BR picture.$^1$ The Coulomb repulsion is thus given by
\begin{eqnarray}
U = P^2U^{\prime} = {\rho}^2U^{\prime}, \\ U^{\prime}={\kappa}U_c,
\\ U={\kappa}{\rho}^2U_c,
\end{eqnarray}
where $0<\rho\le1$, while $0<\kappa{\le}1$ is the correlation
strength. In  the case of ${\kappa}{\neq}$1 and
${\rho}$=1~($\rho_{\uparrow}=\rho_{\downarrow}=\frac12$),   $U$
reduces to the correlation   in the  BR picture. In   the case
when ${\kappa}$=1 and ${\rho}{\neq}$1, $U$ tends   to   $U_c$ when
the band approaches half filling. This indicates that having more
carriers  in  the conduction  band increases the correlation,
$vice ~versa$.

Although $U$ in the Hubbard model is replaced by Eq. (1),
calculations of the expectation value of $U$ based on the
Gutzwiller variational theory$^7$ do not change because $\rho^2$
is constant. Thus we consider the conditions for applying Eq. (3)
to the BR picture. In  the Gutzwiller theory,  ${\bar{\nu}}$ atoms
are  doubly occupied with a probability ${\eta}^{\bar{\nu}}$,
where  $0<{\eta}<1$.$^7$ In the BR picture$^1$ at half filling
${\rho}=1$ and $\rho_{\uparrow}=\rho_{\downarrow}=\frac12$,
${\eta}={\bar{\nu}}/(0.5-{\bar{\nu}})$, and hence
$0<{\bar{\nu}}<1/4$. For  the lowest energy state,
${\bar{\nu}}=(1/4)(1-{\kappa})$ where $U/U_c={\kappa}$,   the
above limits   for ${\bar{\nu}}$ imply $0<{\kappa}<1$. When
${\rho}<1$ and $\rho_{\uparrow}=\rho_{\downarrow}=\frac12\rho$,
the BR picture can be  applied  as  well, with $U$ averaged  over
all sites. Since  the lowest   energy state,
${\bar{\nu}}=(1/4)(1-{\kappa}{\rho}^2)$   where
$U/U_c={\kappa}{\rho}^2$,    is     satisfied    for
$0<{\bar{\nu}}<1/4$, the condition $0<{\kappa}{\rho}^2<1$ is
reached. Thus the inverse of the discontinuity $q$ in the BR
picture,
\begin{eqnarray}
\frac{1}{q}=\frac{m^{\ast}}{m}&=&(1-(\frac{U}{U_c})^2)^{-1},\\
                             &=&(1-{\kappa}^2{\rho}^4)^{-1},
\end{eqnarray}
where $m^{\ast}$  is  the effective mass of a quasiparticle,  is
defined     under the  combined    condition
$0<{\kappa}{\rho}^2<1$, although the  separate condictions are
$0<{\rho}{\le}1$  and $0<{\kappa}{\le}1$.  Therefore, $m^{\ast}$
increases  without bound when ${\kappa}{\rightarrow}$1,
${\rho}{\rightarrow}$1. For ${\kappa}{\ne}$0 and
${\rho}{\rightarrow}$0+, $m^{\ast}$ decreases and, finally,  the
correlation undergoes a (normal or band-type) MIT which differs
from the Mott MIT exhibiting a first order transition. For
${\rho}$=1, Eq. (5) reduces to the effective mass in the BR
picture. At ${\kappa}{\rho}^2=1$, the MIT of the first order
occurs  and the state can be regarded as the   paramagnetic
insulating state because ${\bar{\nu}}=0$. Eq. (5) is illustrated
in Fig. 1 (a).  In addition,  the expectation value of the energy
in the (paramagnetic) ground state is given by
${\langle}H{\rangle}_N={\bar{\epsilon}}(1-{\kappa}{\rho}^2)^2$.
Here, the band energy
$\bar{\epsilon}=\bar{\epsilon}_{\uparrow}+\bar{\epsilon}_{\downarrow}
=2\Sigma_{k<k_F}\epsilon_k<0$ is the average energy
\begin{figure}
\vspace{0.10cm}\centerline{\epsfxsize=8.4cm\epsfbox{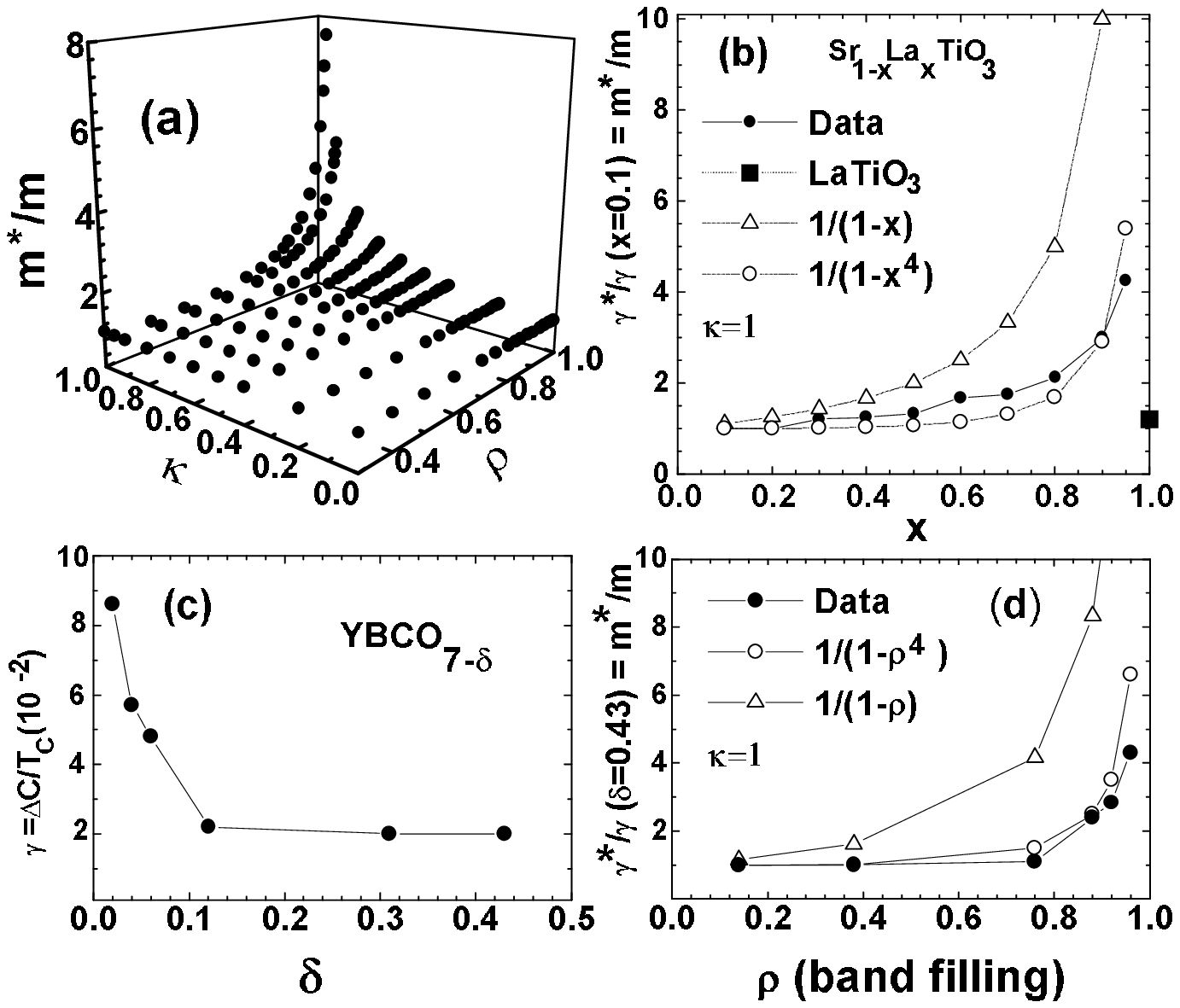}}
\vspace{0.8cm}\caption{(a) The   effective mass,  $\frac
{m^{\ast}}{m}=\frac {1}{1-{\kappa}^2{\rho}^4}$. (b) Experimental
data (${\bullet}$) of the heat capacity  for
Sr$_{1-x}$La$_x$TiO$_3$ presented   by Tokura$^5$    and
Kumagai$^6$,    and  a   comparison   between $\frac
{m^{\ast}}{m}=\frac {1}{1-x}$  calculated from  the $t-J$ model
($\triangle$) and  $\frac {m^{\ast}}{m}=\frac {1}{1-x^4}$
($\circ$).  Here, ${\rho}$ corresponds  to $x$ because  the number
of quasiparticles with    $x$ agrees  with   that of
quasiparticles obtained from the Hall coefficient.$^5$  Here,
${\kappa}$=1 and $0<{\kappa}{\rho}^2<1$. (c) The heat-capacity
coefficient (${\triangle}c/T_c$) obtained by magnetic measurements
for YBa$_2$Cu$_3$O$_{7-\delta}$ superconductors by
Da{\"u}mbling.$^8$ (d) In the cases ${\delta}$=0 and
${\delta}$=0.45, the band is assumed as full with ${\rho}$=1 and
empty with  ${\rho}$=0, respectively. Thus
${\rho}$=1-${\delta}$/0.45. Here, ${\kappa}$=1 and
$0<{\kappa}{\rho}^2<1$.\\}
\end{figure}
 without correlation and $\epsilon_k$ is the kinetic energy in the
Hamiltonian in the BR picture$^1$, with the zero of energy chosen
so that $\Sigma_k\epsilon_k$=0. This picture may be called an
extended BR picture with band filling.

On the other hand, in the $d=\infty$ Hubbard model$^2$, in which
the width of the DOS $\rho(\omega)$ of the coherent part with a
constant peak scale decreases with increasing $U$, the Mott
transition occurs for the ${\it minimum}$ number of
quasiparticles, which
 number is an integral of $\rho(\omega)$ near $U=U_c$ to energies near
$\omega=0$. This is a marked difference from the BR picture in
which the Mott transition occurs at $\rho$=1 (the ${\it maximun}$
number of quasiparticles). The Mott transition in the $d=\infty$
Hubbard model can be regarded as the band-type MIT in the extended
BR picture because of the decrease of the number of quasiparticles
in the coherent part. In addition, the $t-J$ model$^3$ and the
Hubbard model$^4$ on a square lattice predict critical behavior
according to $m^{\ast}/m = 1/(1-x)$.

Eq. (5) is applied to the heat capacity data of the
Sr$_{1-x}$La$_x$TiO$_3$, which is well known as a strongly
correlated system. The number of quasiparticles determined by the
Hall coefficient increases linearly with $x$ up to at least
$x$=0.95.$^{5,6}$ The heat capacity in Fig. 3 of reference 5 is
replotted, as shown in Fig. 1(b). Eq. (5) follows closely the heat
capacity data in  the case when ${\kappa}$=1. The first-order
transition is found between $x$=0.95 and $x$=1. This transition
corresponds to the Mott MIT because ${\kappa}{\rho}^2=1$   in this
picture, with ${\rho}=1$ at LaTiO$_3$ and ${\kappa}=1$ from  the
experimental result. Thus LaTiO$_3$ is a Mott insulator.

Eq. (5) is also applied to the specific heat data for YBCO
measured by magnetic measurements by Da{\"u}mbling$^8$. Eq. (5)
seems to follow the data, as shown in Fig. 1(d). Although  it is
difficult to confirm whether YBCO at ${\rho}$=1 is a Mott
insulator because the effective mass at ${\rho}$=1 is divergent,
the divergence is regarded  as a Mott transition.

${\bf We}$ conclude  from experimental data that the correlation
strength is found to be $\kappa$=1, and that the presently
proposed extended BR picture is better able to explain the Mott
transition than Hubbard models.Further, instead of the van Hove
singularity (vHs), Eq. (5) can be used for 2D-DOS to describe the
mechanism of high-$T_c$ superconductivity.
\\ \\
I thank Mr. Nishio for valuable comments, and Prof. Tokura, Prof.
Kumagai, and Dr. M. Da{\"u}mbling for
 permission to use experimental data in Fig. 1(b),(c).

\end{document}